# Air Gap Control of the Novel Cross (+) Type 4-Pole MAGLEV Carrier System

Enes Mahmut GÖKER
Mechatronics Engineering, Yıldız Technical University, Istanbul, TURKEY
enesgkr@hotmail.com
Ahmet Fevzi BOZKURT
Mechatronics Engineering, Yıldız Technical University, Istanbul, TURKEY
afbozkurt90@gmail.com
Bora BAYKAL
Uskudar American Academy, Istanbul, TURKEY
bbaykal23@my.uaa.k12.tr
Kadir ERKAN
Mechatronics Engineering, Yıldız Technical University, Istanbul, TURKEY
kerkan@yildiz.edu.tr

*Abstract* – Mechanical non-contact carrier systems based on magnetic levitation (MAGLEV) are used in special transportation areas (clean rooms, chemical areas, etc.). Among these types of carriers, 4-pole hybrid electromagnetic systems (containing permanent magnets and electromagnets) stand out with their low energy consumption. The main problem of maglev carrier systems is their non-linear characteristics and unstable open loop response. In this study, PID and I-PD controllers are designed for the air gap control of the new cross-type 4-pole mechanical contactless carrier system.

Thus, the instability problem was overcome and the desired reference tracking for each degree of freedom was successfully carried out in simulation environments, and the results were compared.

**Keywords –** Magnetic Levitation, Maglev Carrier System, PID I-PD Controller Design

## 1 INTRODUCTION

With the developing technology, differences in the technological needs of people have begun to occur. These differences have led to innovations in the fields of production and transportation. Transport of manufactured materials is of great importance. In particular, the material produced in clean rooms, chemical areas and high technology production areas should be away from factors such as vibration, noise and dust. Existing technologies are insufficient for such transport environments (Kim et al. 2011; Jiangheng and Koseki 2001). In the solution of these problems, the use of electromagnetic forces allowed the carrier platform to be levitation, enabling contactless transportation (Atherton, 1980; Han, Kim, 2016; Erkan et al. 2016). The carrier platforms in the literature are divided into two types as active rail and passive rail, but the levitation method is similar in systems where electromagnetic levitation topology is used (Bozkurt et al. 2018; Ertuğrul 2014). U-type, E-type and 4-pole U-type hybrid electromagnets are used for levitation. (Tzeng, Y. and Wang 1998; Lee et al. 2013; Ertuğrul 2014). The proposed cross-type hybrid electromagnet carrier system has a multi-degrees-of-freedom control structure. Each pole produces the electromagnetic force required for magnetic levitation. Pole terminals energized by closed-loop controllers keep the carrier platform levitation and provide the axial or radial movement with internal or external thrust components. Thus, it can be used in multiple engineering applications such as transportation systems, frictionless bearings, and spacecraft design (Jiangheng L and Koseki T 2001).

The cross-type hybrid electromagnet structure contains permanent magnets and electromagnets as in the 4-pole U type electromagnets available in the literature. But due to its structure, it consists of fewer parts compared to the convectional 4-pole. While the convectional 4-pole is formed by the combination of 4 silica sheet metals, the novel cross-type 4-pole is formed by combining two U-types. This has been beneficial for the system to reduce leakage fluxes (Göker E.M. and Erkan K. 2022).

In this article, the model of the novel cross-type 4-pole maglev carrier system has been developed. Since the system has an unstable structure, it needs an active controller. For this, PID and I-PD controllers have been designed by using canonical structure in analytical model and analytical model was linearized, simulation studies have been made. Therefore, by using PID and I-PD type controller, the carrier system can levitate without steady state error. When comparing PID and I-PD controllers for air gap position control, the I-PD controller provides the best results.

## 2 MAIN CONTENT

### 2.1 Levitation Model of Cross Type Hybrid Electromagnet

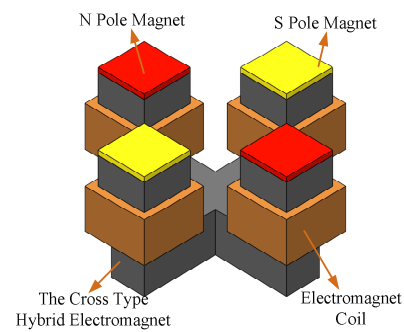

Figure 1. The cross-type hybrid electromagnet

The core and coil windings of the cross-pole hybrid electromagnet are shown in Figure 1.

The dimensions of the cross-type 4-pole hybrid maglev carrier system and the centralized control geometric transformation matrices are given (Göker E.M. and Erkan K 2022).

Controllers are designed on this model by linearizing the system in the predicted operating range. Ferromagnetic body resistance, magnetic saturation, hysteresis fuco losses and flux leakages are neglected in the modelling.

The operating point where the linearization is made is the point where the axes of rotation are absent. If balancing is done at this point;

$$(z, \alpha, \beta) = (z_0, 0, 0)$$
$$(i_z, i_\alpha, i_\beta) = (i_{z_0}, 0, 0) \quad (1)$$

should be. The dynamics of movement in the Z-axis direction at this point;

$$m\Delta\ddot{z}(t) = f_z(z(t), i_z(t)) - mg - F_d(t) \quad (2)$$

If $f_z$, which we express as the gravitational force,

$$f_z(z(t), i_z(t)) = \frac{B^2}{2\mu_0} \cdot S \cdot 4 = \frac{2\mu_0 S(E_{PM} + Ni_z(t))^2}{(z(t) + l_{PM})^2} \quad (3)$$

is expressed as. $E_{PM}$ denotes the AT value of permanent magnets, N stands for winding count, $i$ stands for pole current, $l_{PM}$ magnet thickness, S stands for magnet area, $\mu_0$ stands for magnetic permeability constant. In Figure 2, the air gap and the force graph formed by the system against the currents are given.

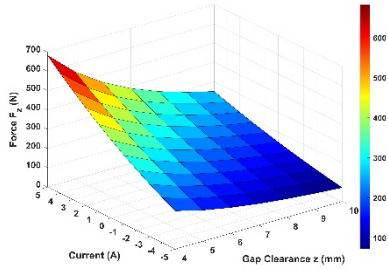

Figure 2. Electromagnetic Attractive Force Characteristic in Z Axis

The linear model for the levitation system of the electromagnet is found by linearizing the electromagnetic attraction force by choosing as in the 1rd equation. Minor variations around the linearized model of the system;

$$f_z(z(t), i_z(t)) = f(z_0 - \Delta z(t), i_{z_0} + \Delta i_z(t))$$
$$= f_z(z_0, i_{z_0}) + K_A\Delta z(t) + K_B\Delta i_z(t) \quad (4)$$

expression is made. Here;

$$K_A = -\frac{\partial f_z}{\partial z}\bigg|_{(z_0, i_{z_0})} = \frac{4\mu_0 S(E_{PM} + Ni_{z_0})^2}{(z_0 + l_{PM})^3} \quad (5)$$

$$K_B = \frac{\partial f_z}{\partial i_z}\bigg|_{(z_0, i_{z_0})} = \frac{4\mu_0 SN(E_{PM} + Ni_{z_0})}{(z_0 + l_{PM})^2} \quad (6)$$

formula is obtained. If Equation 2 is rearranged;

$$m\Delta\ddot{z}(t) = K_A\Delta z(t) + K_B\Delta i_z(t) - F_d(t) \quad (7)$$

obtained. m; mass (kg), $F_d$; is expressed as the disturbance input force (N). If the laplace transform of Equation 7 is done according to the air gap;

$$Z(s) = \frac{K_B}{ms^2 - K_A}I_z(s) - \frac{1}{ms^2 - K_A}F_d(s) \quad (8)$$

available in the form. Angular displacement axis dynamics;

$$J_\alpha \frac{d^2\theta(t)}{dt^2} = T_\alpha(\theta(t), i_\alpha(t)) \quad (9)$$

$$T_\alpha(\theta(t), i_\alpha(t)) \cong T_\alpha(z_0, \theta(t), 0, i_{z_0}, i_\alpha(t), 0) \cong K_C\Delta\theta(t) - K_D\Delta i_\alpha(t)$$

$$K_C = \lim_{\theta \to 0} \frac{\partial T_\alpha(\theta, 0)}{\partial \theta}$$
$$K_D = \lim_{i_\alpha \to 0} \frac{\partial}{\partial i_\alpha}\left\{\lim_{\theta \to 0} T_\alpha(\theta, i_\alpha)\right\} \quad (10)$$

is expressed as. Since the cross-pole maglev carrier has a symmetrical structure, the β axis dynamics shows the same characteristics as the α axis dynamics. For this reason, the β-axis dynamic equations are not included.

In Equation 8, the control signal is given in the form of a current source. The coils used in levitation system can be energized by using a voltage source instead of a current source. From this point of view, if the dynamic equations are expressed again;

$$e(t) = R_z i_z(t) + N\frac{d}{dt}\Phi = R_z i_z(t) + \mu_0 NS\frac{d}{dt}\frac{Ni_z(t) + E_{PM}}{z(t) + l_{PM}} \quad (11)$$

is expressed as and this equation is linearized;

$$\Delta e_z(t) = R_z \Delta i_z(t) + L_z \Delta \dot{i}_z(t) + \frac{K_A L_z}{K_B}\Delta \dot{z}(t) \quad (12)$$

form is obtained. Taking the laplace transform according to the air gap of the equation;

$$I_z(s) = \frac{1}{L_z s - R_z}\left[E_z(s) - \frac{K_A L_z}{K_B}sZ(s)\right] \quad (13)$$

is found. The linearized system dynamics for the Z-axis is given in Figure 3 with block diagrams.

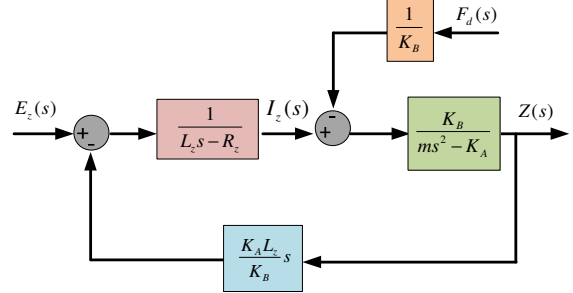

Figure 3. Z-Axis Linearized System Dynamics

In Table 1, the model parameters of the cross-pole electromagnet are extracted.

Table 1. Cross Pole Electromagnet Model Parameters

| Size/Unit | Value | Size/Unit | Value | Size/Unit | Value |
|---|---|---|---|---|---|
| m [kg] | 10.00 | $z_0$ [mm] | 19.30 | $\alpha_0, \beta_0$ [rad] | 0.00 |
| $J\alpha,\beta$ [kg.m²] | 0.30 | $i_{z0}$ [A] | 0.00 | $i_{\alpha 0}, i_{\beta 0}$ [A] | 0.00 |
| S [A²] | 12*10⁻⁴ | $K_A$ [N/m] | 12473 | $K_C$ [Nm/rd] | 79.28 |
| β [AT/N] | 19.85 | $K_B$ [N/A] | 9.88 | $K_D$ [Nm/A] | 2.38 |
| $R_z, \alpha, \beta[\Omega]$ | 1.00 | $L_{z,\alpha,\beta}$ [H] | 0.016 | $E_{pm}$ [AT] | 3970 |

## 2.2 Simulations and Results

There are 4 control inputs to control the levitation of the cross-pole hybrid electromagnet. These control inputs can be modelled in 3 axes (z and α, β axis) by transforming them into central control axis matrices. The Z axis model is similar to the α, β axis models, and the difference only appears in the relevant parameters. The model of the voltage input system is given in Equation 14.

$$Z(s) = \frac{K_B}{(R_z + L_z s)(ms^2 - K_A) + K_A L_z s} E_z(s) - \frac{R_z + L_z}{(R_z + L_z s)(ms^2 - K_A) + K_A L_z s} F_d(s) \quad (14)$$

### 2.2.1 PID Controller Design

PID controllers are the most used controllers in the literature. PID controllers are used in many fields due to their simple structure, low number of control variables and easy physical implementation. The PID structure of the z-axis controller of the cross-pole hybrid electromagnet is given in Figure 4.

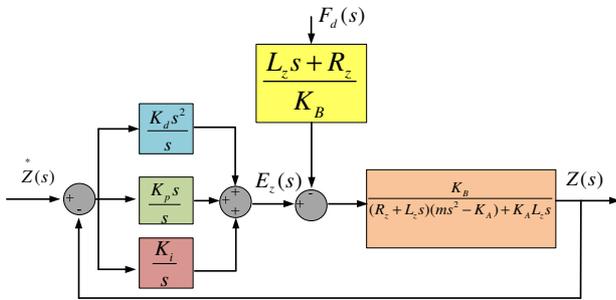

Figure 4. Structure of The System's z Axis PID Controller

When the transfer function of the block diagram is found;

$$\frac{G(s)}{1+G(s)} = \frac{Z(s)}{Z^*(s)} = \frac{K_B(K_d s^2 + K_P s + K_i)}{K_B(K_d s^2 + K_P s + K_i) + s[R_z + L_z s(ms^2 - K_A) + K_A L_z s]} \quad (15)$$

obtained. The canonical polynomial approach was used to calculate the controller coefficients in the closed-loop transfer function (Mochizuki and Ichihara 2013; S. Manabe. 1998).

Table 3. PID and I-PD Controller Coefficients

| Kp | Kd | Ki |
|---|---|---|
| 1756 | 32 | 3088 |

The Matlab-Simulink model is given in Figure 5. In this simulation, the response of the system against the entered air gap reference value is observed.

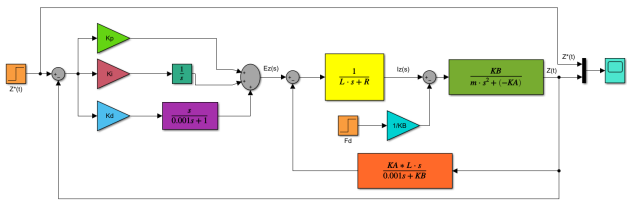

Figure 5. PID Controller Simulink Model

Looking at Figure 6, it is seen that the PID controller creates an overshoot by exceeding 1.5 millimetres against the 0.5 mm reference signal given in the 1st second. In the real system, this overshoot causes the system to become instability (mechanical constraints and non-linear characterism). This problem can be solved by using the I-PD controller instead of the PID controller.

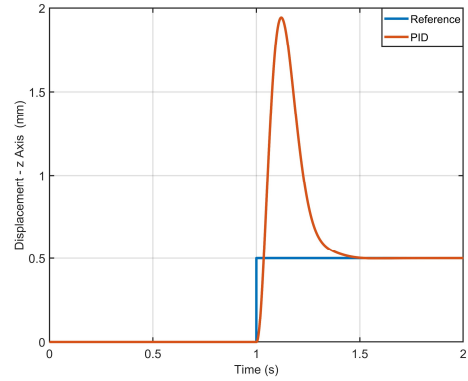

Figure 6. Step Reference Input PID Controller z Axis Response

### 2.2.2 I-PD Controller Design

Significant changes in the reference signal input in the PID controller produce large input signals as a result of proportional (P) and derivative (D) coefficients. This causes overshoots in the system, that is, saturation. Since the integral (I) block is used first in the I-PD controller, it integrates the first effect on the error signal. Thus, it limits the input signal to be applied to the system. Proportional and derivative expressions are integrated into the system as feedback from the closed loop. (Mochizuki and Ichihara 2013).

When the transfer function of the block diagram is found;

$$\frac{Z(s)}{Z^*(s)} = \frac{K_B K_i}{s^4(mL_z) + s^3(mR_z) + s^2(K_B K_d) + s(K_P K_B - K_A R_z) + K_B K_i} \quad (16)$$

is expressed.

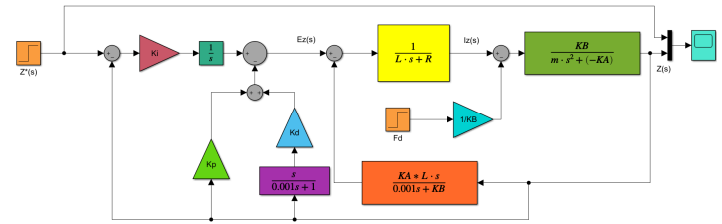

Figure 7. I-PD Controller Simulink Model

Figure 7 shows the simulation study of the I-PD controller. In this simulation study, a reference air gap value was entered and the system was controlled in this air gap with the I-PD controller.

The model was carried out using the parameters used in PID. Looking at Figure 8, it is seen that the I-PD controller responds more smoothly to the 0.5 mm reference signal given at 1 second, without creating an overshoot.

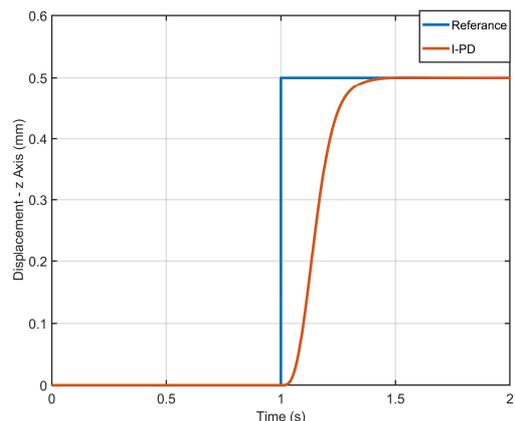

Figure 8. Step Reference Input I-PD Controller z Axis Response

Figure 9 shows the comparison of PID and I-PD controllers. Looking at this comparison, it is seen that using the I-PD controller instead of the PID controller is better in terms of control.

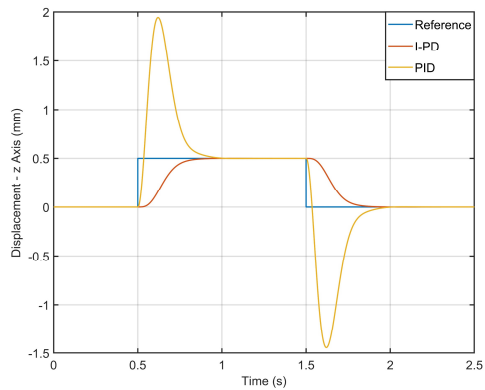

Figure 9. Comparison of Step Reference Input I-PD Controller with PID Controller's z Axis Response

## 3 CONCLUSION

In this study, the air gap control of the maglev carrier system, which contains a novel cross-shaped 4 pole hybrid electromagnet, was carried out. Analytical model was linearized, PID and I-PD controllers were designed by using canonical structure in the analytical model and controllers were compared in simulation environment. The successful realization of simulation provides opportunities for future studies.

For future studies, it is aimed to perform the experimental setup and to improve the system performance by studying different controllers.